\documentclass[sigconf]{acmart} 
\AtBeginDocument{%
  \providecommand\BibTeX{{%
    \normalfont B\kern-0.5em{\scshape i\kern-0.25em b}\kern-0.8em\TeX}}}

\usepackage{csquotes}
\DeclareUnicodeCharacter{200B}{{\hskip 0pt}}
\setcopyright{acmlicensed}
\copyrightyear{2024}
\acmPrice{15.00}
\acmDOI{10.1145/1122445.1122456}

\acmConference[CHI '24]{Proceedings of the 2024
CHI Conference on Human Factors in Computing Systems}{May 11--16, 2024}{Hawai'i, USA}
\acmBooktitle{Proceedings of the 2024 CHI Conference on Human Factors in
Computing Systems (CHI '24), May 11--16, 2024, Hawai'i, USA} 
\acmPrice{15.00}
\begin{document}

%\newcommand{\highlightchange}[1]{{\color{black}{ #1}}}

\title{\#PoetsOfInstagram: Navigating The Practices And Challenges Of Novice Poets On Instagram}

\author{Ankolika De}
\email{apd5873@psu.edu}
\orcid{}
\affiliation{%
  \institution{College of Information Sciences and Technology, Pennsylvania State University}
  \country{USA}
}

\author{Zhicong Lu}
\email{zhicong.lu@cityu.edu.hk}
\orcid{}
\affiliation{%
  \institution{Department of Computer Science, \\City University of Hong Kong}
  \country{Hong Kong SAR}
}
% \affiliation{%

% \affiliation{%

% \affiliation{%

%\renewcommand{\shortauthors}{XXX, et al.}
%\newcommand{\zl}[1]{{\textcolor{blue}{ [Zhicong: #1]}}}

\begin{abstract}
Commencing as a photo-sharing platform, Instagram has since become multifaceted, accommodating diverse art forms, with poetry emerging as a prominent one. However, the academic understanding of Instagram's poetry community is limited, yet its significance emerges from its distinctive utilization of a primarily visual social media platform guided by recommendation algorithms for disseminating poetry, further characterized by a predominantly novice creative population. We employ qualitative analysis to explore motivations, experiences, and algorithmic influence within Instagram's poetry community. We demonstrate that participants prioritize conforming to algorithmic constraints for visibility, yet maintain their community's values of integrity and originality, illustrating the tension between algorithmic growth and participant authenticity. We introduce the concept of \textit{Algorithmically Mediated Creative Labor}, a phenomenon specific to non-monetizing creative users who are impacted by the prioritization of professional creators and continually adapt their creative endeavours to align with platform logic, thereby affecting their motivation and creative outputs.
\end{abstract}






\maketitle

% \emph{For Internal Use Only}
\section{Introduction}

Social media platforms have enabled non-professional users to actively engage in content creation and sharing, resulting in a substantial expansion of the content creator base \cite{simpson2023, jenkins2006fans, jose2009}. This phenomenon has diversified content genres within these platforms, fostering a surge of creative expressions. The rise of the Instagram poetry community is a consequence of innovative utilization of the platform to propagate traditional art forms \cite{novakovic2021instagram, jeong2016hanbok, poetryisnotdead}. The accomplishments of influential Instagram poets like Rupi Kaur, boasting 4.5 million followers, and Atticus, with 1.6 million followers, further underscore the community's exponential growth trajectory, which is consistently reinforced by the continual influx of novices \cite{atlantic}. Within this community, individuals use Instagram as a dynamic platform for poetry, fostering sub-communities marked by mutual interactions that support their members \cite{poetryisnotdead, novakovic2021instagram}.

Human-Computer Interaction (HCI) research has extensively investigated creator communities situated within online platforms, aiming to illuminate their operational dynamics \cite{kou2017, chi2021, kelly2002, eric2010}, algorithmic underpinnings \cite{youtubealgo, 10.1145/3555149, simpson2023}, and creative labor \cite{simpson2023, brooke2021sm, 10.1145/3596671.3597655}. Notably, a predominant focus has been on content creators of a professionalized nature, individuals who intend to commercialize their artistic endeavors \cite{chi2021, 10.1145/3555149, youtubealgo}. Consequently, these inquiries have explored creators who possess a dual proficiency—both in their chosen artistic domain and in navigating the intricacies of online platforms \cite{chi2021, Cotter, 10.1145/3555149, devitoann2022}. Although prior research has explored the activities of hobbyist creators, the emphasis has predominantly revolved around understanding these creators' comprehension of the algorithms \cite{youtubealgo}. Notably, there is a gap in the literature concerning the labor and specific work practices affecting novice creators who do not aspire for professionalization. The predominant existence of novices engaging in creative activities online is particularly due to the widespread accessibility of these platforms. This accessibility enables a diverse group of users to actively participate in content creation, contributing to the generation of \textit{user-generated content} (UGC) \cite{jose2009}. Therefore, delving into the creative activities of novices not only sheds light on the evolving nature and labor that goes into online participation but also emphasizes the broader impact of UGC in shaping digital landscapes. 

The poetry community on Instagram is characterized by a multitude of novices who leverage a platform predominantly designed for visual content to share a distinctly textual art form, appropriating its features in an unconventional manner \cite{hennessey2015extrinsic, Langello_2014}. Given the dearth of scholarly attention directed towards the poetry community on Instagram within academic discourse, coupled with the existing gap in examining the experiences of novice creatives with the algorithm, particularly in comprehending the labor involved, our objective is to elucidate the values, motivations, and prevalent practices within this community. Consequently, we aim to examine how individuals understand and interact with algorithms particularly highlighting the labor and extra effort novices face when working within platforms influenced by algorithms. We ask: 

\begin{itemize}
\item \textbf{RQ1: Why do novice poets share their poetry on Instagram? What are the values and practices of the poetry community on Instagram?}

\item \textbf{RQ2: How do the novice poets of Instagram perceive the recommendation algorithms of Instagram, and how does this perception impact their practices?}
\end{itemize}

To address these questions, we conducted in-depth, semi-structured interviews with novice poets on Instagram (N=15) and observed Instagram posts (n=240) with popular hashtags within the poetry community. The first author, adopting a complete-member researcher role \cite{adler1987membership}, utilized personal connections for recruitment and drew on her own experiences for analyses, elaborated further in Section \ref{researchpos}. Drawing from a grounded theory-inspired approach \cite{charmaz2017constructivist}, we uncovered themes regarding the values, practices, and algorithmic perceptions of the poetry community on Instagram.

We found that participants are unable to \textit{authentically} post and share their poetry, instead having to give in to algorithmic constraints to achieve any \textit{visibility} on the platform. We further highlighted the culture and practices within the community that distinguished them from other creators, particularly showing their participatory values, wherein as a community they strove to maintain \textit{integrity} and \textit{originality}. We advocate for similar studies on novice, non-professionalized creative communities and emphasize platform support for creators without monetary motivations, especially by drawing from Simpson and Semaan's \cite{simpson2023} redefinition of \textit{creative labor}. Our design suggestions emphasize authenticity, encompassing transparent algorithms and community-based recommendations, especially beneficial for emerging creatives. Additionally, we propose equitable recommendation algorithms, where novice creatives are prioritized to foster a supportive digital environment. We conclude with the juxtaposition adhering to algorithmic preferences for growth and visibility, despite its potential impact on participant authenticity and creative labor.

We contribute to the HCI community in two major ways. Firstly, we explicate the values and practices prevalent among novice creators by analyzing the perspectives of poets within the Instagram poetry community, thereby highlighting the platform's unique role as a creative space for many emerging writers. Secondly, we introduce the concept of \textit{Algorithmically Mediated Creative Labor}, focusing on novice creators who struggle with creativity due to inadequate platform support for authentic sharing and growth.

\section{The Poetry Community of Instagram: A Background}
Poets, historically, have been recognized as highly collaborative individuals, often engaging in literary circles, and community gatherings to share ideas and improve their craft collectively \cite{fisher2003open, daniels2002literature, golden2000use}. In the context of online communities and social media platforms, prior research has demonstrated the effectiveness of collaborative creativity, leading to successful outcomes in various artistic domains \cite{10.1145/2470654.2466266, stockleben2017towards}. The collaborative nature of poets on Instagram serves as an intriguing subject, particularly considering how this approach empowers both experienced poets and novices alike \cite{golden2000use, waldie2021collaborative}.

Despite the growing presence and influence of Instagram poets within contemporary literary circles, their significance has been largely overlooked in traditional academic discourse. Paquet, (2019) reported that this neglect was largely due to their reputation of being \textit{uncultured} and \textit{not serious}, a stance that is surprising since Instagram poetry refurbishes poetry as an art form at a time when it is fading \cite{lili2019}. The lack of attention and recognition \cite{lili2019, poetryisnotdead} given to Instagram poets within traditional academic circles signifies a hesitancy or resistance to assimilate creative works originating from digital platforms, such as social media. The struggle to accommodate these new digital art forms indicates the need for a more inclusive scholarly approach. Thus, our study, characterized by the democratization of literature through Instagram poetry \cite{lili2019, manning2020crafting}, highlights the transformative power of technology in fostering unique creative communities and empowering emerging individual artists. Our work hopes to provide multidisciplinary perspectives on the entanglements of creativity and technology.

\section{Related Work}

% creativity, authenticity and originality: 
% Creative, emotional and visibility labor in algorithmic mediated systems.

%why these RQs?

%online communities formed by content creators
%online creative communities

In this section, we perform an extensive literature review, emphasizing content creators' significant role in social media's growth and online community development. We also discuss research on creativity and authenticity, focusing on creators' self-presentation and social media's affordances. Finally, we examine the labor involved in online content sharing and its impact on creativity.

\subsection{Content Creators and Online Communities}

We draw from extensive HCI research on content creators on social media (e.g., \cite{chi2021, choi2023, youtubealgo, Sharma2022ItsAB, camilla2022, cheng2022, devitoann2022, ma2022}) to investigate poets of Instagram. Turri et al., \cite{Turri2013DevelopingAB} defined content creators as users who actively participated on digital platforms through interactions or by sharing their own content. Subsequent inquiry has involved categorizing users into distinct groups: \textit{inactives}, \textit{spectators}, and \textit{joiners} representing those who engage passively without significant content contribution. Additionally, \textit{critics}, \textit{conversationalists}, and \textit{creators} have been identified as users actively involved in content generation \cite{li2011}. This classification gained notable support, especially in light of the extensive adoption of UGC \cite{10.1145/1358628.1358692, 10.1145/2470654.2470659}. UGC aids in platform development, often originating from non-professional creators who contribute to its growth \cite{10.1145/1358628.1358692, 10.1145/2470654.2470659}. Additionally, UGC is almost always creative and spread the idea and practice of content creation amidst non-professional users \cite{Mileros2019TowardsAT}. 
    
Our work situates the poets as \textit{creators} who regularly share their work and \textit{conversationalists} that use platform features to engage in discussions with other poets and passive users. Beyond the fact that they have largely been ignored by the academic community, we are interested to study this community of creators for three main reasons. Firstly, we are interested because this group primarily comprises of novice creatives \cite{kiernan2021instagram}. Contrary to the limited prominent figures within the community who have attained global recognition, the majority of individuals assume a novice status, engaging in poetry primarily as a leisure pursuit \cite{kiernan2021instagram}. Their utilization of Instagram is predominantly driven by a desire for artistic expression rather than a pursuit of fame, providing them with a platform to channel their creative endeavors \cite{vadde2017amateur}. Secondly, Instagram poets stand apart from other creators on the platform due to their distinct utilization of the platform's visual-centric environment for textual expression \cite{camilla2022}. While Instagram predominantly caters to visual media sharing, such as images and videos \cite{rogers2021}, these poets leverage the platform's features to showcase written compositions. This unique multimodality of predominantly textual content \cite{lili2019} within a visual context sets them apart. Finally, Instagram poets engage with poetry, an age-old art form, adapting it to a contemporary digital medium, thereby creating a distinctive intersection between tradition and modernity \cite{martens2021portrait}. This unique approach not only challenges the platform's dominant content paradigm but also offers an innovative channel for creative expression, contributing to the platform's diversity.

HCI scholars have also been interested in unraveling the intricate dynamics of online communities (e.g., \cite{dosono2019, ahmadi2020, preece2004, 10.1145/1753326.1753658, 10.1145/1518701.1518848, 10.1145/3170427.3173032}), primarily due to the influence of content creators who shape and nurture these digital collective spaces \cite{ringland2022}. Scholars are motivated to comprehend how individuals engage in digital spaces and how technological interfaces shape these interactions. This pursuit has yielded insights into user engagement \cite{malinen2015understanding, guo2021, o2022rethinking, lampe2010}, information dissemination \cite{lueg2001information, eva2018}, social dynamics \cite{nicolas2006, gero2023}, and interface design \cite{Turri2013DevelopingAB, kougray}, contributing to the formulation of user-centric design principles for digital platforms' usability and effectiveness \cite{stonefr, zhengux}. Within this landscape, the poets of Instagram present an intriguing avenue of study, given their distinct role in community formation and interaction patterns \cite{poetryisnotdead}. Understanding how they navigate collaborative spaces, leverage visual elements, and manage the constraints of the platform can further scholarly understandings of the evolving landscape of digital poetry and its impact on creative expression. Consequently, we hope to contribute to the broader HCI discourse on diverse content creators and their interactions within digital ecosystems.

\subsection{Creativity and Authenticity}

HCI researchers have extensively explored creativity, particularly its support through technology (e.g., \cite{10.1145/3461778.3462050, 10.1145/3386526, 10.1145/3475799, tang1991videodraw, 10.1145/570907.570939, choi2023, simpson2023, gero2023}). We draw from Plucker et al.'s description of creativity--- an interaction amidst thought, process and surroundings that is individual or collaborative, leading to the forming of something that is of relevance in a particular context \cite{pluck2004}. Supporting this definition, we also follow that broader socio-cultural contexts exert a substantial influence on the shaping and impact of creative expression \cite{amabile2012}. 

Poetry serves as a vital artistic medium enabling individuals to engage in \textit{comparative analyses} \cite{simecek2016uses}, constituting dimensions of politics, metaphor, and substantial pedagogical value. This significance extends beyond collective audiences to encompass personal introspection \cite{parini2008poetry, faulkner2017poetry}. Likewise, teaching and learning poetry has been an important part of the curriculum in educational institutions. Researchers have found that poetry is learnt best when writers interact with social contexts that are most relevant to them \cite{dymoke2017}. In accordance with this perspective, technologically proficient younger generations employ Instagram for pedagogical poetry sharing \cite{poetryisnotdead}.

Digital platforms also shape artistic expression in different ways \cite{das2022, tyler2022}. For instance, they can act as authentic learning environments that are important for learning and practicing creative arts \cite{Engelman2017CreativityIA}. Past studies have found that digital platforms allow individuals who are often marginalized and censored in real life to authentically present themselves online, behind controlled anonymity, thus empowering members of such communities to organize and support one another \cite{duguay2019running, duffy2019gendered}. Likewise, the proliferation of creators and creativity on digital platforms has been studied extensively, especially since they afford social creativity, with features particularly focused towards creative collaborations \cite{yue2018, cheng2022}. 

While these platforms offer a channel for creative expression, their impact is multifaceted, encompassing both facilitation and constraint. Algorithmic logic, at times, inadvertently curtails the scope for genuine innovation \cite{tanvir5unraveling}. Amidst this backdrop, our investigation gravitates toward self-presentation, an avenue of exploration that has garnered notable scholarly attention. Creatives, often leverage their artistic productions for self-presentation, since their work is a culmination of their selves \cite{eikhof2019creativity}, and filtering to meet the constraints of social media, can hinder genuine authenticity. The concept of authenticity consistently punctuates this discourse, signifying a recurring theme that underscores the significance of genuine artistic expressions \cite{articleathelete, devitohow}. The ongoing dialogue between self-presentation strategies and algorithmic hindrances highlights how creators grapple with maintaining authentic creative pathways amidst the platform constraints \cite{choi2023, manning2020crafting}. This discourse, which intertwines technology, self-presentation dynamics, and authenticity, not only underscores the importance of our investigation but also resonates with a broader landscape of research into the intricacies of digital creativity.

Poets, constituting a diverse group of creators, employ language as their artistic medium, skillfully negotiating complexities of self-presentation, cultural dynamics, and multifaceted authenticity \cite{poetryisnotdead, lili2019, kleppe2018poetry}. By exploring their creative processes and interactions with digital platforms, our research offers a distinctive perspective for investigating the dynamics of how authenticity and creativity manifest in contemporary sociotechnical contexts, particularly within the previously unexamined realm of diverse content creators.

%add in 2.1

\subsection{Creative, Emotional and Visibility Labor in Algorithmic Mediated Systems}

The labor involved in interacting with algorithms on social media platforms is a significant aspect of modern digital engagement \cite{rosenblat2016algorithmic}. Users exert cognitive effort and strategic planning to adapt to algorithmic cues, to enhance their content's visibility \cite{kang2022ai, brett2019}, reflecting the dynamic interplay between human agency and algorithmic influence and underscoring the evolving nature of user interactions within digital environments. These algorithms, designed to tailor content recommendations based on platform goals, necessitate users' active involvement in crafting and curating their online experiences \cite{bhandari2022s, wu2019}. Additionally, it requires users to adapt their content to align with algorithmic preferences. This process underscores the relationship between human intent and algorithmic influence, shaping user interactions on social media \cite{wu2019, devitoann2022, devitohow}.

The infrastructure of digital platforms forces creatives to actually \textit{be creative} in a manner that their work fits into and is visible within \textit{platform logic} \cite{Abidin2016VisibilityLE, brooke2021sm, choi2023} --- referring to prevailing conventions, strategic approaches, processes, and economic foundations that form the basis of the platform's functionality \cite{simpson2023, 2013van}. Consequently, creators must adapt to and create content, such that they can be understood and translated amidst platform logic \cite{ma2021}. These processes encompass labor that extends beyond the mere act of content creation, to ensure content alignment with and integration into the operational framework of the platform \cite{Abidin2016VisibilityLE, brooke2021sm, choi2023}. This integration includes not only adherence to algorithms but also extends to broader platform logic and dynamics, including but not limited to platform profitability, management, and interface changes \cite{Abidin2016VisibilityLE, brooke2021sm, choi2023}.

Additionally, algorithmic workings also introduce an element of precarity, characterized by the unpredictability of performance metrics when sharing content \cite{brooke2021sm}. Creatives' actions to meet platform metrics and conform to their norms demonstrate how creators are enmeshed in the broader systems of platform capitalism (a complex economic framework revolving around digital platforms that facilitate a range of exchanges, labor arrangements, and economic undertakings \cite{srnicek2017platform, liang208, srnicek2017platform}). Furthermore, sharing content on digital platforms means that creatives have to show \textit{calculated authenticity}, where they have to consistently ensure that while they share authentic work, they limit their content to what is appropriate in each context \cite{duffy2017mythologies}. Such issues additionally increase the emotional labor of creatives. 

While the poetry community of Instagram has not been seen to have members of broadly underrepresented populations, it is known to have individuals who produce deep, personal work that is difficult to homogenize within platform infrastructures \cite{lili2019, selfpres1, poetryisnotdead, faulkner2017poetry}. Instagram and analogous platforms are recognized for their personalization attributes, predominantly driven by their recommendation algorithms \cite{parmelee2023personalization, mahnke2017ripinstagram}. However, these algorithms have been criticized for creating filter bubbles characterized by repetitive and discerning content \cite{10.1145/3336191.3371825}. Such concerns have been voiced by content creators in prior research \cite{herm2021, Fischer2018IdentifyingAE, semenzin2022swipe}. Creativity has been known to suffer in such environments, as without exposure to new ideas, new ways of thinking are difficult to develop \cite{reviglio2017serendipity, pariser2011filter, onitiu2022fashion}. 
 
Recent scholarly discourse has spotlighted the intricate nature of creative labor within the context of digital platforms \cite{simpson2023}. The definitional boundaries of creative labor have been subject to scrutiny, a consequence of the unique affordances these platforms offer while simultaneously entailing laborious efforts to align innovative content within their frameworks \cite{simpson2023}. The evolving understanding of creative labor recognizes that while platforms grant diverse opportunities for self-expression, their rigid logics can necessitate substantial labor for attaining congruence between creative vision and platform-specific functionality \cite{simpson2023, smith2015m, hong2020you, devitohow}. Additionally, a broader concern pertains to the specter of homogenization \cite{chaney2018, caplan2018isomorphism}, wherein the distinctive attributes of creative work could be overshadowed by standardization, underpinning the significance of redefining and optimizing creative labor. This pertains not only to creators seeking monetary gains but equally to those with non-monetary aspirations \cite{simpson2023}. Studying the poetry community assumes particular importance, propelled by the prevalence of novice poets within its ranks \cite{kiernan2021instagram}. This community serves as a distinctive lens for investigating creative labor within digital platforms because it sheds light on the distinctive characteristics of labor within emerging creative communities.

\section{Methods}

\begin{table*}[t]
\centering
\small
  \caption{Self-Reported Participant Demographics.}
  \label{tab:freq}
  \begin{tabular}{ccccccccl}
    \toprule
   ID & Gender & \# of Instagram Followers & \# of Instagram Following & Nationality & Age & Last Obtained Education\\
    \midrule
    
    P01 &Female &2463 &1101 &Indian &20 &High School Degree\\
 P02 &Female &2961 &1691 &Indian &23 &Bachelors Degree\\
 P03 &Female &533 &258 &American &19 &N.A.\\
 P04 &Female &2704 &1442 &Indian &19 &High School Degree\\
 P05 &Gender Fluid &1609 &1030 &Indian &26 &Post Graduate Degree\\
 P06 &Female &266 &166 &American &23 & Bachelors Degree\\
 P07 &Male &1899 &527 &Indian &24 &Bachelors Degree\\
 P08 &Female &1006 &312 &Indian &24 &Bachelors Degree\\
 P09 &Male &217 &126 &Indian &19 &High School Degree\\
 P10 &Male &990 &196 &Indian &25 &Bachelors Degree\\
 P11 &Non Binary &374 &164 &French &21 &High School Degree\\
 P12 &Female &578 &402 &American &21 &Bachelors Degree\\
 P13 &Male &4795 &650 &Indian &26 &Post Graduate Degree\\
 P14 &Female &718 &379 &Indian &23 &Bachelors Degree\\
 P15 &Female &4613 &753 &Indian &24 &Post Graduate Degree\\

  \bottomrule
\end{tabular}
\end{table*}

% \subsection{Methodology}

Due to the highly contextual \cite{howarth2016contextual} and exploratory \cite{stevens2013exploratory} nature of our study, we used qualitative methods, including semi-structured interviews (N=15) and content analysis of posts (N=240) to address our research questions. Instagram was used to both recruit participants and collect posts for observations. Drawing from grounded theory \cite{charmaz2017constructivist}, we used inductive thematic analysis to highlight relevant findings that answer our research questions \cite{kyngas2020inductive}.

\subsection{Data Collection}
We employed two distinct data sources in our study, namely Instagram content (posts) and interview transcripts.

\subsubsection{Instagram Posts}
Between May and September 2022, the first author, leveraging her own knowledge and other literature on the poetry community \cite{poetryisnotdead, lili2019}, utilized commonly recognized hashtags such as \#poetry, \#poetryisnotdead, \#instagrampoetry, and \#poemsofinstagram, to conduct a targeted search for relevant posts. The top, 300 posts per hashtag were selected. Subsequently, a manual curation process was employed to refine the dataset. Posts that did not exhibit a clear connection to poetry, such as those lacking written poems or content that deviated significantly from poetic themes, or those not in English (given that the authors possessed proficiency solely in that particular language) were systematically removed. After the initial filtering step, a random selection of 60 posts per hashtag was introduced to minimize algorithmic bias, ensuring a diverse sample that included content beyond the top-ranked posts in our analysis. This curation process created a focused dataset, comprising of 240 posts, all of which were captured and stored as screenshots to maintain their original content and context. 

\begin{table}
\footnotesize
  \caption{Themes of Writing in Instagram Posts from Content Analysis}
  \label{tab:themes}
  \begin{tabular}{ccccccccl}
    \toprule
   Theme & Frequency\\
    \midrule
    Heartbreak& 97\\
    Grief& 33\\
    Other social issues& 25\\
    Love& 20\\
    Self-help& 19\\
    Politics& 17\\
    Religion& 17\\
    Care& 12\\

  \bottomrule
\end{tabular}
\end{table}

\begin{table}
\footnotesize
  \caption{Length of Poetry: In our sample, we observed that the majority of posts were brief, as indicated by the word count. This trend aligns with our research findings, where participants noted that shorter posts tended to receive greater amplification from the algorithm.}
  \label{tab:char}
  \begin{tabular}{ccccccccl}
    \toprule
   Length & Frequency\\
    \midrule
    
    <500 Characters& 193\\
    =>500 Characters& 47\\

  \bottomrule
\end{tabular}
\end{table}

\begin{table}
\footnotesize
  \caption{Relationship between visual and textual materials: Here, the first author read through the content that contributed as poetry and checked the visual for the same. If there was any relationship, i.e., even one line of the poem was associated with the visual, it was coded \textit{yes}.}
  \label{tab:relationship}
  \begin{tabular}{ccccccccl}
    \toprule
   Are the texts and pictures related? & Frequency\\
    \midrule
    
    Yes& 39\\
    No& 201\\

  \bottomrule
\end{tabular}
\end{table}

\subsubsection{Participant Recruitment}
Recruitment was carried out through a combination of snowball and purposeful sampling methodologies \cite{Robinson2014SamplingII}. The first author used direct messaging on Instagram to engage with contemporary peers, who were poets. Concurrently, these poets were encouraged to refer other interested individuals within their network. Furthermore, a recruitment advertisement was disseminated via an Instagram story. All interactions conducted through direct messages were followed by requests for participants' email addresses, which served as a channel for subsequent communication. Participants with under 5000 Instagram followers were recruited to maintain consistency in expertise and usage duration. The lead author used purposeful sampling to select Instagram users with under 5000 followers and assessed profiles found through snowballing for compliance with this criterion, to ensure that this study remains grounded in novice experiences.

Individuals who expressed interest were further engaged through email correspondence, involving the distribution of a screening survey, that asked if they self-identified as Instagram poets. Subsequently, a comprehensive consent form was provided, elucidating participants' rights and the voluntary nature of their participation in the study. Finally, interviews were scheduled, with reaffirmation of consent obtained verbally at the commencement of each interview session. Following the interviews, we administered an online survey to gather demographic data.

Participants, had a mean age of 23.4, and follower counts ranging from 217 to 4795. All participants wrote poetry in English. The majority (9 out of 15) were women, consistent with prior studies in poetry \cite{bangart}. Among them, three were Americans, one French, and the rest Indians (See table \ref{tab:freq}). 

\subsubsection{Interviews}

Our interviews, conducted between August and December 2022, were semi-structured and voluntary. They aimed to explore poets' motivations, interactions, and perceptions of the algorithm. The interviews began with the first author sharing her own experiences as an Instagram poet, to maintain rapport and emphasize on shared ideas and goals \cite{Drabble2016ConductingQI}. As participants grew more comfortable with sharing, the interviewer asked questions particularly regarding their own ideas and motivations for using the platform. An example of this sequence, was the interviewer starting with how she started sharing on the platform, after a particularly bad day in her life, to which the participant shared that on such days, "\textit{the poetry community was their only hope}". On building a rapport based on shared experiences, the said participant went on to discuss other experiences and motivations that they had with the community. We then asked them about the platform, its affordances and their particular practices. Additionally, we asked them about the algorithm, especially discussing ways in which they perceived it, and its impact on their practices.  

Recruitment concluded upon achieving data saturation, as defined by Charmaz and Belgrave (2012), indicating the point at which no substantially novel information emerged during the interviews \cite{charmaz2012qualitative}. This determination was made through concurrent preliminary analysis processes, ensuring that adequate information had been gathered to comprehensively address the research questions. 

The interviews, conducted virtually due to global distances, occurred via Zoom, were recorded, and transcribed for analysis. Ranging from 30 to 57 minutes, with a mean duration of 45 minutes, the interviews were conducted in English, and participants were thanked upon completion.

\subsection{Data Analysis}
Our research methodology, was drawn from constructivist grounded theory \cite{charmaz2017constructivist}, and employed an inductive-iterative based thematic analysis approach \cite{article04, qual}. Thus, while we did not use grounded theory, we used many concepts from the same, particularly its iterative coding process \cite{charmaz2017constructivist} when pursuing thematic analysis. 

We began with an analysis of our transcripts, and honed our interview protocol as themes were generated to ensure that our findings remain grounded in the data \cite{charmaz2017constructivist}. Our approach to transcript analysis adhered to an iterative thematic analysis methodology \cite{article04, qual}. The first phase encompassed three rounds of open coding, where we coded the transcripts line by line \cite{charmaz2017constructivist}, generating codes that showed \textit{fit and relevance} \cite{charmaz2017constructivist} for our research questions. Examples of these codes included \textit{emotional feedback} and \textit{selling out for fame}. Subsequently, we engaged in two rounds of focused coding \cite{charmaz2017constructivist} to synthesize information from these initial codes, identifying overarching themes such as \textit{motivations} and \textit{algorithmic perception}, that particularly corresponded to our research questions.

Simultaneously, we performed qualitative content analysis \cite{mayring2019qualitative, charmaz2017constructivist} on the Instagram post data, guided by our interview data. We mainly coded for visual styles and elements within the posts, investigating themes, visuals, and lengths of posts (see ~\autoref{tab:themes} and ~\autoref{tab:char}) as those were the main topics of discussions in our interviews. Additionally, we examined the relationship between visual and textual components in the posts, since participants explained that to be confusing, and often impacted by their algorithmic perception (see \autoref{tab:relationship}). Content analysis supplemented our interviews by offering concrete examples and context of the values and practices discussed by the participants. Furthermore, as it occurred simultaneously with the interview analysis, we also tailored our protocol based on insights gleaned from Instagram posts and the distinctive practices we observed. For example, while the initial interview protocol did not extensively cover specific themes, we identified distinct patterns during the content analysis, particularly in relation to the disparities between the visual content and accompanying texts. Recognizing the significance of these themes and such differences, we proactively modified the interview protocol to incorporate targeted questions aimed at understanding the reasons behind the observed discrepancies. This adjustment allowed for a more nuanced exploration of the factors influencing participants' behavioral choices on Instagram, enriching the interview process with a more contextual interplay between images and text in their posts.

The first author was involved in the actual coding process, while the second author provided an outside perspective to further relate the themes with the interview data. 

\section{Research Positionality Statement}
\label{researchpos}

The first author was a complete-member type researcher \cite{adler1987membership}, meaning that she was already an existing member of the poetry community of Instagram. Her struggles with Instagram's algorithmic inconsistencies, discovered through interactions with fellow novice poets, led her to formulate the research questions addressed in this study. As part of the poetry community, she orchestrated the data collection, leveraging her own network for purposeful sampling and extending reach through snowball sampling. During interviews, she fostered trust by assuring participants their experiences would not be portrayed negatively, establishing mutual empathy as a fellow community member. In this way, the interviewees were honest and open about their experiences \cite{adler1987membership}, and could speak about their algorithmic experiences and perceptions freely, without much concern about being judged.

The second author was an outside member, and provided a different perspective, particularly in their understanding of the said community directly from the data. This distinction in experiences between the authors, was helpful for maintaining objectivity in the research, while also representing the needs and challenges of the participants accurately. Consequently, the analysis was informed and characterized by the relative positionality of the authors to their relationship with the topics of discussion \cite{charmaz2017constructivist}.

\section{Findings}

In this section, we report our findings across two sections, that broadly correspond to our research questions. In the first section, we report motivations, values, and practices prevalent in the poetry community, and specifically investigate poets' perceptions of the algorithm, along with its impact on their work. We find that poets are unhappy with the algorithm, especially since as novice creatives, they are unable to navigate it efficiently. The next section describes the specific challenges that poets encounter with the platform. In particular, we focus on their status as non-monetizing creatives that the algorithm inadvertently sidelines compared to more mainstream content creators. 

\subsection{Motivations, Values, and Practices of the Poetry Community of Instagram}

The poetry community on Instagram engages in unique practices, some of which align with other content creators, while others cater specifically to the needs of poets. We report that while Instagram offers a convenient platform for sharing content, with various affordances to elicit creativity and sharing, it falls short of fully grasping and appreciating the complexities and depth of creative work that extend beyond visual aesthetics. Additionally, we discuss poets' perception of the recommendation algorithm, and extend it to report interesting insights on how their work is shaped by platform logic. This section raises questions about how platforms like Instagram can better cater to a diverse range of creative forms and provide a more holistic experience for artists. 

\subsubsection{Motivations for using Instagram}

Our research aligns with prior studies on Instagram creatives, indicating poets' inclination towards community, networking, and collaboration. Furthermore, we identified unique motivators, such as instant validation, routine development, asserting readership authority (especially among hesitant novice poets), and leveraging Instagram's user-friendly interface, all essential in driving poets' creative pursuits.

\textbf{Instant Feedback}.
12 out of 15 participants shared that the validation that they receive within the Instagram community is a significant motivator for their writing. P09 noted, "\textit{Instagram has been instrumental in motivating my consistent four-year poetry writing journey with its encouraging feedback and support}." This validation loop serves as a driving force for sustained creativity. P10 supported this, "\textit{Comments on my post, such as "I love the second line" and "The meaning resonated with me and inspired my writing," motivated me to continue despite almost giving up.}"

Numerous (n=11) participants spoke about how accountability is fostered through the visibility offered by the platform. P06 compared her poetry account to her account where she documented her running, "\textit{My running account taught me consistency. So, I started using Instagram for regular poetry writing, which I hadn't done before and it worked!}" P03 elaborated on this, explaining that shared creative journeys lead to unity and accountability within the community, stating, "\textit{It is a logistic tool to keep up the regularity of writing, promoting some amount of self discipline among us}."

\textbf{Control Over Readership}.
Several participants (9 out of 15) explained that Instagram affords them a controlled readership environment, enabling them to share their work openly without the constraints of traditional publishing. P08 expressed, "\textit{Hard copy publishing is not something I am ready for, and I can post on Instagram without any judgement and eventually publish something more}," while P12 affirmed, "\textit{I can control who reads my poetry and if I want to block someone or archive something at any point, I can do that too! It gives me the freedom to write more openly.}"

\textbf{Ease of Usage}.
13 out of 15 participants claimed that the user-friendly interface and multifunctional capabilities of Instagram are key drivers for their work. 11 poets lauded the platform's efficiency in consolidating posting, feedback, and social interaction on a single platform. P01 remarked, "\textit{You can post poetry, give feedback, receive comments and socialise on the same platform}," while P13 concurred, "\textit{Despite its issues, IG provides a convenient all-in-one platform for sharing poetry, unlike blogs or other stuff.}"

\subsubsection{Unique Values and Practices of the Poetry Community}

Our participants explained their distinct needs, values and practices, with P05 succinctly sharing,

\begin{quote}
    "\textit{The poetry community is about heartfelt expression, not money or fame. We share our art, learn, and support each other. Plagiarism is also flagged, and good work should be amplified. It's a great community!}" (P05)
\end{quote}

The poets had interesting values and practices, that separated them from many mainstream creatives. In this section, we report the poetry community's distinctive values, including their participatory sharing, emphasis on originality, and their use of textual content on a visual social media.

\textbf{Participatory Sharing}.
Within tight knit groups, members of the poetry community engaged in vibrant exchanges, sharing insights, strategies, and advice that helped navigating the challenges of a predominantly visual social media landscape. P08 shared, "\textit{If I do something and the algorithm particularly amplifies my post because of it, I quickly share my tactics with others in my circle.}" Similarly, P05 shared, "\textit{I engage with fellow new poets, sharing work and growth tips. Limited followers mean we actively promote our work for visibility}." Likewise, 7 out of 15 poets agreed that simply sharing posts, never allowed any visibility on their work. Instead, participating in proactive re-sharing and direct dissemination of posts using stories and direct messaging was essential to gain more like-minded audiences.

\textbf{Importance of Originality}.
A significant aspect of the poets' unique practices revolved around the paramount importance placed on originality. This community prioritized originality, encouraging members to infuse their unique voices and perspectives into their poetry. Some (n=11) poets mentioned that while going through their \textit{explore page} or recommended posts, they look for work that is original, authentic, and different. P02 shared on this topic, 

\begin{quote}
    "\textit{I want to read poetry that is new and authentic. I want to see what part of the writer's life is depicted in the poems, and I want to understand the diverse group of people that form this community. I am not interested in stuff that is repeated, or work that is simply an extension of whatever is popular at the moment.}" (P02)
\end{quote}

Our participants concurred that while profit-oriented creators leaned toward \textit{popular content}, the poetry community sought authentic recommendations. This perspective distinguishes poets from other creatives due to their emphasis on seeking newness and authenticity in poetry. They express a desire to connect with the writer's personal experiences reflected in the poems and to gain insights into the diverse community of poets. This underscores a distinct creative approach from trend-focused content creators. P09 shared in this context, 

\begin{quote}
    "\textit{Us poets are old school. We have been meeting in groups for ages. We share our unique poems and get inspired by each other's beauty. But if we're always stuck creating the same stuff, it's tough to stay passionate and keep writing.}" (P09)
\end{quote}

This commitment to individuality sets them apart from other creative groups, emphasizing their dedication to producing original content that genuinely captures their distinct creative essence.

\textbf{\textit{Selling Out} for Visibility}. 11 out of 15 participants noted several people in the community to be \textit{sellouts}, wherein their work was simply a superficial piece on whatever was most trending on the platform. P04 shared on this, "\textit{Posts with just a few words about trending topics get thousands of likes. I don't see how real voices can succeed in a platform focused on engagement.}"

Participants (n=11) shared, that such poets who attempt to change their content to gain wider attention, can be perceived negatively by others within the community. Such individuals face accusations of \textit{selling out} or compromising their artistic integrity. This dynamic underscores the community's commitment to preserving the essence of their craft and resisting commercial influences.

\textbf{Upholding Values as a Community: The Case of \textit{Plagiarism}}. Amidst the poetry community's work, instances of plagiarism have emerged as a noteworthy concern. To address this challenge, the community of poets have developed a proactive approach. When suspicions of plagiarism arise, poets frequently leverage the \textit{stories} feature on Instagram to illustrate the parallels and distinctions between original compositions and the potentially plagiarized content. This shared vigilance underscores their dedication to preserving authenticity and upholding integrity.

9 poets contended that due to the unregulated nature of this creative community, coupled with the absence of formal review processes, occurrences of plagiarism have become more prevalent. P10 elucidated this perspective, stating, 

\begin{quote}
    "\textit{I see similar or copied posts, try to find the original source, and ask others for help. If the person doesn't remove the copied content after we ask, we share it on our stories to spread the message.} (P10)"
\end{quote}

In essence, the poetry community's unique practices on Instagram are a reflection of their predominantly novice status. Their participatory culture, emphasis on originality, and their proactive stance against plagiarism contribute to an intricate tapestry of values that shape their work on Instagram.

\subsubsection{Algorithmic Superficiality.}
\label{algsup}

Our participants deemed the recommendation algorithm superficial, lacking a deeper understanding of their work. P06 shared-

\begin{quote}
    "\textit{I love Instagram, I love that it's so easy to use, and everything- all the templates are already built. But Instagram doesn't understand me, its not for creatives like us. It pretends to. I think it just gets pictures, its doing what I would call, judging the book by its cover in the sense that it judges the poems with the picture, and this harms us, because what do we do? Should we follow our heart or should we please the algorithm?}" (P06)
\end{quote}

P06's assertion that Instagram \textit{pretends to} understand creative individuals highlights the platform's poor attempt to accommodate various content types, including textual content like poetry. However, this perceived understanding is incomplete and focused primarily on visual elements. The analogy of \textit{judging the book by its cover} underscores the platform's tendency to make surface-level assessments based on the accompanying image rather than delving into the deeper layers of meaning and emotion conveyed by the textual content itself. This often leads to poets posting pictures that are unrelated to their actual poems, leading to "\textit{another layer of complexity and confusion}" (P13) in decision making during their sharing process. Our content analysis also showed, how most poems that the algorithm promoted under the hashtags had visuals and texts that did not match (See table 4). 

According to P07, "\textit{I think Instagram knows what kind of photos I like, but when it comes to poems, it still thinks it's the photos I care about.}" P05 further emphasizes this by stating, "\textit{I am constantly shown poems of poets who just share glamorous pictures as opposed to the deeper stuff that is more abstract that I actually like.}" These observations underscore the algorithm's inclination towards prioritizing visual content over the deeper thematic layers intended in poetry, even when poets aim for more abstract or personal textual material.

These insights show the algorithm's limitations in understanding nuanced content. Its proficiency lies in recognizing explicit visual cues, such as colors or patterns, rather than fully comprehending the intricate nuances of textual content, especially within poetry. This deficiency becomes particularly apparent for participants (n=11), whose work often draws from socio-economic intricacies and employs less direct language than other content creators. 

Finally, a common theme among all 15 participants is the tension between artistic authenticity and algorithmic conformity, causing them to unintentionally deviate from their original creative intentions to cater to the algorithm's preferences. P01 shares here, 

\begin{quote}
    "\textit{Sometimes, I just unintentionally post a picture of me with a poem about social justice, because at the back of my head I know it will get more reach if I do that. But, then I realize how stupid that looks, and I go back and want to change it, and ofcourse then my post literally has like 5 likes}"
\end{quote}

P07 captured this struggle, noting that it takes effort to \textit{unlearn the algorithm} and resist altering their work to meet its expectations.

\subsubsection{Personal Nature of the Poets' Work}

The poets believed that the personal nature of their work, made it difficult to survive on Instagram. In order to actually produce such work, they required complete freedom to at least produce the content, and not be tied to algorithmic constraints to ensure that their work is seen and discovered. In this context, P03 shared, "\textit{I love writing, sharing, and reading poetry, but the whole visibility and shadow-ban worry kills my vibe. Sometimes I just stop halfway and wonder if it's even worth it.}" They began their social media journeys by completely depending on platform structures to amplify their work, however they realised that without constraining to certain platform conformities their motivations of receiving feedback, validation and meeting other like minded poets, were not achieved which hampered their creativity and produced a lack of accountability. 

Poets also relied on the recommendation system to help discover authentic work. However, algorithmic recommendations often led to posts primarily featuring visually appealing images with brief statements on topics like love or heartbreak (see Table \ref{tab:themes}), lacking the personal depth they valued. P04 and P05 shared the sentiment that most content that they discovered via the recommendation system were "\textit{lacking depth}" and "\textit{simply famous because it was someone's picture with barely any text.}" This undermined interviewees' motivation to share on Instagram, which centered around discovering content and learning from new connections. The increasing content homogenization led poets to independently search for their preferred work, further exacerbated by a lack of personalized creative content search tools on Instagram. P01 shared, "\textit{I manually search on Instagram, but searching within a community or specific post subset isn't possible.}"

\subsection{Challenges of Algorithmic Mediation for Novice Creatives}

Our interviewees were novices in creative pursuits. Despite being engaged in poetry writing for an average of five years, they had only recently (average: 1.5 years) commenced sharing their work on social media platforms. With increasing utilization of Instagram, they encountered prominent poetry accounts, which in turn fueled their aspirations for account growth.

\subsubsection{Algorithmic Hindrance towards Growth}
All our interviewees believed that the platform, and its algorithms were responsible for the lack of visibility of their posts and their slow growth. Their beliefs came from two major observations. As mentioned in section \ref{algsup} they found that the recommendation algorithms that drove content, were trained on more visual and multi-modal forms of content, and amidst that competition, Instagram and most other social media platforms were inefficient in being used as a poetry sharing medium. P02 shared,

\begin{quote}
    "\textit{Instagram understands the pictures. So if its a cat, but my poem is about how we should be more like cats, come and go as we please, I find my poem in a cat-lovers page completely out of the poetry community? Like Instagram literally judges the book by its cover!}"
\end{quote}

Similarly, P07 shared, 

\begin{quote}
    "\textit{I used to think, more attractive pictures were what drove the algorithm, so I started putting pictures that had no relation to the poem, and it kinda makes me sad but otherwise IG just does not share my stuff!}" 
\end{quote}

Our content analysis also showed that most posts had pictures and poems, where the visuals did not actually match the text (see table \ref{tab:relationship}), complementing the interview data, that explained how this way the algorithm only notices visually attractive content and amplifies it more.

Secondly, poets (n=12) felt that the algorithm only helped those that used its monetizing features or worked unfaithfully towards increasing their engagement in growth. This belief stemmed from their observations of peers that had also started out with them, but quickly shifted to commercial modes of growth, with creativity and original sharing taking a secondary seat. P05 shared in this context, 

\begin{quote}
    "\textit{I see how accounts and especially some of the people I had met while starting to share on Instagram started working with ads and had the goal of monetizing. They started sharing more attractive posts and quickly grew. I don't understand how to actually grow as a poet without also making money on it.}" 

\end{quote}

P07 added, 

\begin{quote}
    "\textit{I really hate trying to give into monetizing factors while writing poetry. I would really like to just write, share and have the algorithm spread my stuff. Why does it always have to be about an ulterior motive?}"
\end{quote}

Additionally, our content analysis also showed that the majority of posts had less than 500 characters, showing that most poets that the hashtags actually amplified were short, thus hindering longer, more nuanced poetry and promoting shorter, more entertaining content (See table \ref{tab:char}). 

\subsubsection{Algorithmic Isolation of Non-monetizing Creatives}

Participants shared that if they did not use the monetizing features on Instagram, the algorithm did not support their content. 

\begin{quote}
    "\textit{Instagram only supports you if you are trying to make money off it, and in turn giving it money. What about us? Our content enriches the platform, it diversifies and helps train its recommendation algorithms! Why are we not supported?}"(P06)
\end{quote}

Many (n=10) believed, that the \textit{algorithm isolated} individuals seeking a space for personal growth and creative refinement, without placing a central emphasis on monetary gains. P05 spoke about this lack of support, 

\begin{quote}
    "\textit{I think Instagram only has features that help creatives who want to make money out of their work. We are learning and we like doing poetry! But I, nor a lot of others want to actually make a career out of it!}" 
\end{quote}

Participants (n=6) underscored that Instagram's framework tends to prioritize activities that yield economic benefits, inadvertently sidelining those poets who join the platform with a primary intention of fostering their creative growth. P03 specifically shared, 

\begin{quote}
    "\textit{I hate that they have all those insight tabs and creative accounts to help creators make money. I want to know how to \textbf{not} use ads but still grow!}" 
\end{quote}

Likewise, P04, shared, "\textit{I work, write poetry for fun. I want fair visibility for my posts without constant changes or relying on the unpredictable algorithm.}"

Participants stress the need for a more inclusive Instagram support system that caters to diverse poetic goals, particularly emphasizing creative growth over financial gain.

\subsubsection{Algorithmic Hindrance Towards Authenticity}

12 poets shared that achieving favorable outcomes necessitated adhering to specific thematic and stylistic trends, as well as visual elements, to enhance visibility within the platform's algorithms. P11 shared, "\textit{How do I get new followers when Instagram only amplifies people who already have a million followers.}" As a result, poets often found themselves compelled to conform their creative expressions to topics that didn't naturally align with their work. P05 shared in this context, "\textit{I do five posts and then on the sixth one, I generally incorporate some stuff that I find is trending, otherwise none of my work gets seen by anyone!}" 

As mentioned in section \ref{algsup}, in their urge to ensure that their work gets discovered by the algorithm, and consequently by other poets, they often had to derail from their actual work. To do so, they often changed their themes of writing to align with the algorithm's preferences, diluting their authentic voice. Some poets revealed that when their work didn't conform to favored themes, it held deeper \textit{personal significance} but received little recognition, leaving them demotivated.

\subsubsection{Algorithmic Favoring of Themes}
The themes that mostly drove the algorithms, were apparently created by what participants (n=14) called \textit{poet influencers}, referring to those poets that had large fan followings. P03 shared, 

\begin{quote}
    "\textit{You've seen famous poets on social media, right? They write about popular topics and get tons of likes, but I want to address fairness for farmers. The problem is, these influencers have already set the trend, and my topic isn't favored by the system.}"

\end{quote}

Our participants mentioned themes such as grief (n= 4), love (n= 7) and sex (n= 4) among others as their top choices. From our content analysis, we found similar \textit{personal} themes of writing (see table \ref{tab:themes}). Participants saw poetry as a deeply personal form of expression. While they could learn practical tips for using the platform from \textit{more famous poets} (P02, P04, P07, P09, P10-P13), limiting themselves to themes favored by popular creators hindered their creativity and overall creative process. P02 shared here, "\textit{I'm inspired by Instagram poetry giants, but this realm is personal. I don't want to follow trends; I value writing from my heart, not just popular themes.}"

Conforming to particular work themes without having any actual knowledge of what the algorithm wanted was also difficult. P12 shared, "\textit{I might write about love because a similar post got 3000 likes, but even then, the unpredictable algorithm's response is uncertain.}"

\subsubsection{Algorithm-Induced Degradation in Quality}
13 out of 15 poets felt that larger poetry accounts were losing quality, as they grew since they conformed to trending themes, and lost all sense of originality. P04, shared,

\begin{quote}
    "\textit{I see larger accounts, and sometimes their work is just trash. They just write a few random lines about love and call it poetry and then the algorithm amplifies it, and other posts that follow similar themes. But it is not poetry! Its trash!}"
\end{quote}

This made the poets (n=13) feel that if they pertained to said norms, their work too would lose quality. Algorithmic preference for seemingly less sophisticated poems garnering significant engagement introduces a distinct challenge. Poets navigate a dilemma: the tension between authentic expression of intricate themes and adherence to algorithmic trends for visibility. This highlights the distinctive complexity within the poetry community, where the interplay between artistic authenticity and algorithmic conformity is particularly pronounced.

\section{Discussion}
%signpost

Our findings highlight the poetry community's distinctiveness in seeking immediate feedback and desiring reader control. We also noted unique community values, emphasizing their participatory culture and commitment to preventing plagiarism. Finally, we emphasized the significance of authenticity and originality in the poetry community, with participants sharing challenges posed by recommendation algorithms hindering their creative development.

In this section, we contextualize the unique needs and challenges of the poetry community. We also highlight potential repercussions if these challenges are not addressed. Additionally, we introduce the concept of \textit{Algorithmically Mediated Creative Labor}, as the work that it takes for non-monetizing creatives to navigate algorithms while striving to be creative, ultimately hindering their goals and demotivating their work.

\subsection{\#poetsofinstagram: A Unique Perspective of Novice Content Creators}

\begin{figure*}[t]
    %\centering
    \includegraphics[width=0.8\linewidth]{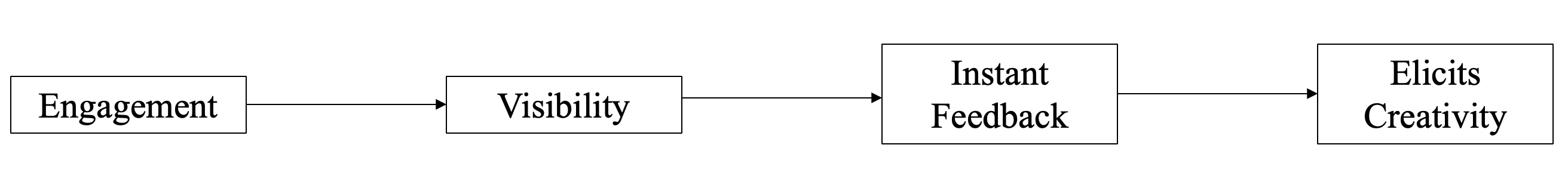}
    \caption{Poets need to gain engagement to remain creative. Unless they work with and navigate the algorithms efficiently, they will not be able to tap into the affordance of instant feedback, thus reducing accountability and creativity.}
    \label{fig:pge}
\end{figure*}

Analyzing Instagram poetry in the UGC landscape reveals the evolving role of novice creators, utilizing the platform as a canvas for authentic expression, particularly extending literature on novice content creators on social media. By prioritizing authenticity, these novice poets exemplify a trend within the larger UGC community, where personal expression takes precedence over established norms. The challenges outlined, such as those tied to platform logic and visibility requirements, can resonate with the broader experiences of novice creatives engaging in UGC, offering a glimpse into the diverse ways in which individuals leverage social media platforms for creative expression without necessarily seeking professionalization.

Participatory culture plays a central role in the Instagram poetry community, fostering a dynamic environment for creativity to thrive. Poets actively engage with one another, offering feedback and support, thereby enhancing the development of their craft. Previous research has demonstrated that instant feedback plays a pivotal role in stimulating creativity \cite{eric2010, kyung2007}. This insight is particularly significant for diverse online creatives, including poets, and contribute to their artistic development \cite{10.1145/2804405}, while also extending to informal knowledge sharing practices among online communities \cite{pandey2016pop}. However, to leverage the affordance of instant feedback on Instagram, poets are reliant on their work being rendered visible by the platform's algorithm. Consequently, poets face a dual challenge: not only must they create content that resonates with their own and intended audiences' values, but they also need to navigate the platform's logic to ensure their work is amplified by the recommendation systems (See \autoref{fig:pge}). The tensions between authenticity and visibility can be disruptive to a creative process by hindering the \textit{state of flow} \cite{getzels1976creative} that most creatives, including poets require while engaged in their work. \textit{Flow} refers to how, when one is involved in creative work an intrinsic motivation that is independent of any quantitative output drives the individual through creative work \cite{getzels1976creative}. On prioritizing growth, leading to conforming to practices that support visibility, this state of flow can be diminished, leading to decreased creativity. 

Additionally, participatory values within the community, such as publicly highlighting instances of plagiarism and copying, contribute to fostering an atmosphere of integrity and respect. Drawing from our findings, the pursuit of visibility and growth on the platform inadvertently leads certain poets to imitate popular styles or themes with higher engagement rates, resulting in them being stigmatized as \textit{sellouts} while others admitting it as being the sole means of expanding their account's reach. From a media equation theory lens \cite{reeves1996media}, this dilemma can result in poets and other creatives who struggle with similar tensions think of algorithms as gatekeepers \cite{youtubealgo, napoli2015social}, an authority figure, shaping their visibility and exposure leading to creating content that would appease and conform to platform logic, since they perceive their content's visibility as an endorsement of its value and quality. 

Furthermore, the algorithmic reinforcement of popular content can perpetuate a feedback loop \cite{mansoury2020feedback}, where certain styles or themes dominate the platform, making it challenging for less conventional or niche poetry to gain exposure, leading to homogenization of content \cite{nechushtai2023more, nechushtai2019kind}. This can be disheartening for poets with unique perspectives, inhibiting their willingness to share their work and engage in the community.

In conclusion, novice creatives such as Instagram poets \cite{jose2009, simatzkin2022user}, significantly contribute to generating UGC, enhancing social media platforms. Social media platforms capitalize on the extensive volumes of UGC data they accumulate, utilizing it as a key element in their primary monetization strategy. This data plays a pivotal role in the creation of targeted advertisements and personalized recommendations, thereby contributing to the overall success of the platform \cite{Andrejevic2009ExploitingYC}. Recognizing and promoting the involvement of diverse creators in shaping UGC is thus crucial for maintaining the dynamic and inclusive characteristics of platforms. By neglecting support for novice creatives who produce UGC, platforms impede marginalized voices and suppress content diversity \cite{doi:10.1177/0163443711427199}. In contrast to earlier UGC discussions focused on marketing \cite{doi:10.10520/EJC-173db9a0b4} and trust in online content \cite{cheong2008consumers}, our research identifies and explores a community that actively engages with and enhances UGC, fostering artistic growth and development among its members. We underscore the importance of supporting free creatives, especially novice poets, to enhance content diversity on platforms and foster creative pursuits.

\subsection{Algorithmically Mediated Creative Labor}

\begin{figure*}[t]
    \centering
    \includegraphics[width=0.8\linewidth]{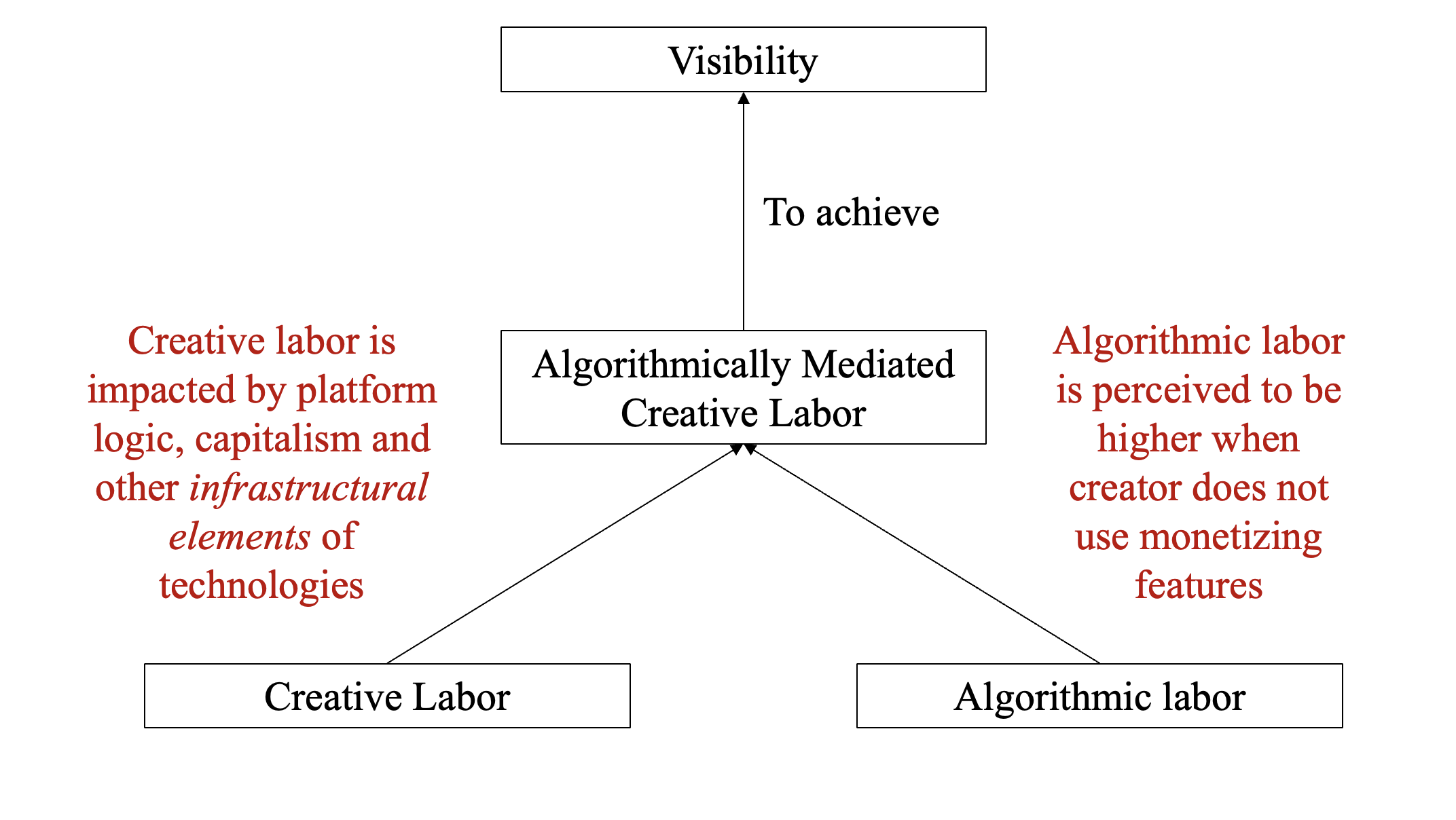}
    \caption{Algorithmically Mediated Creative Labor: The conflicting demands between algorithmic conformity and authenticity create a state of confusion and labor for novice/ non-monetizing creators, necessitating an intricate balance. Consequently, creators find themselves expending significant efforts in both creative and algorithmic domains, striving to generate artistic content while adhering to algorithmic constraints. Nevertheless, engaging in this demanding process proves essential, as it underpins the visibility and recognition of their creative endeavors.
}
    \label{fig:acml}
\end{figure*}

The tension between appeasing the algorithmic preferences of Instagram and staying authentic increases the creative labor for poets within the platform. We draw from recent literature on redefining \textit{creative labor} including \textit{how the routine enactment of creative labor is impacted by infrastructural elements of technology} \cite{simpson2023} to specifically investigate how the poetry community, composed of mainly novice, non-professionalised content creators can be supported by digital platforms such as Instagram. The sociotechnical definition underscores the implications of creative labor for non-monetizing novice poets, considering its multifaceted dimensions lacking mere monetization and professionalization efforts. We argue that the labor that it takes to be creative in platform environments is algorithmically mediated. We define \textit{Algorithmically Mediated Creative Labor} (AMCL) as the exertion of effort by creatives, particularly novices, in efficiently navigating platforms' monetizing bias to pursue their creative endeavors (Refer to \autoref{fig:acml}). We discuss the implications of AMCL in the subsequent sections. 

\subsubsection{Loss of Creative Autonomy} We discovered that the tension arising from AMCL had a significant impact on the motivation and discipline of poets within the Instagram poetry community. As poets sought to strike a balance between conforming to the platform's algorithmic preferences and staying true to their authentic creative selves, many reported experiencing a notable decline in their motivation to write and maintain their creative output. Supported by self-determination theory (SDT) \cite{deci1980self}, this can lead to the loss of intrinsic motivation for creative expression among diverse creators. As the pressure to appease the algorithm takes precedence, the satisfaction derived from the creative process itself (intrinsic motivation), which drives the creative flow \cite{getzels1976creative} is overshadowed by external factors, like the desire for likes and shares (extrinsic motivation). Drawing from prior work by Klotz et al., (2012), \cite{klotz2012can} this shift can lead to a decreased sense of autonomy and competence, ultimately impacting poets' and other creatives' motivation to write and share their work \cite{klotz2012can}.

The decline in motivation and discipline to write also resulted from the frustration of seeing algorithmically favored but insincere content receive more attention than genuine poems, as shared by P04 who explicitly called out the platform and \textit{sellouts} for hindering creativity. The perceived prioritization of profit-driven content over artistic merit led many poets to question the integrity of the creative community and the platform itself, further dampening their enthusiasm to contribute. Additionally, the tension created a sense of unpredictability in the outcomes of their creative efforts. Creatives found it challenging to gauge how their work would fare algorithmically, as the platform's preferences seemed to constantly change \cite{ma2021}. Beyond the poetry community, this has implications in  discouraging all novice creatives from maintaining a consistent creative output while also fostering a feeling of being at the mercy of external forces beyond their control \cite{brooke2021sm}.

The advent of AMCL has the potential to significantly impact novice creatives, raising concerns about its effects on the authenticity and diversity of creative expression \cite{simpson2023, tanvir5unraveling, frich2018hci}. In this study, we argue for the prioritization of free creatives on digital platforms, as they form the cornerstone of \textit{creative} UGC ecosystems \cite{10.1145/1358628.1358692}. Neglecting their contributions may impede the growth of novices and hinder non-monetizing individuals from engaging in creative endeavors, potentially leading to stagnation in platform innovation.

\subsubsection{Sustainable Innovation and Growth}

Modern digital platforms heavily rely on UGC to drive user engagement and maintain their competitive edge \cite{tanvir5unraveling}. Amidst this landscape, free creatives such as novice poets and other novice creators, who contribute without direct monetary incentives, play a crucial role in generating diverse and dynamic content \cite{tanvir5unraveling}. However, the prevalence of AMCL has introduced challenges for these creatives. The pressure to align their content with algorithmic preferences, along with a lack of support without intentions of monetizing can lead to a loss of creative autonomy and an inclination towards predictable content, as is supported by our findings, where participants complained of homogenization in creative communities. Thus, we advocate for platforms to prioritize the needs of free creatives to foster an environment that sustains the \textit{continued growth} of UGC ecosystems.

We argue that supporting growth such that their creative autonomy is preserved is vital, in maintaining the flow of fresh and diverse content, otherwise leading to stagnation in UGC ecosystems \cite{ekbia2017heteromation}. Platforms that encourage independent creative expression can break away from algorithmic-induced homogeneity \cite{chaney2018, ekbia2017heteromation}, allowing for a broad spectrum of content that resonates with users. By embracing diversity, encouraging originality and innovation, and nurturing a sense of community, platforms can ensure continuous growth, enhance user experiences \cite{shepherd2012persona}, and safeguard the dynamic nature of their own systems and infrastructures in the face of algorithmic influences \cite{jose2009, john2017, subramaniam2021}.

\subsection{Desgin Implications to Support Poetry Sharing and Other Forms of Creativity}

The primary aim of this study is to promote the creativity of the Instagram poetry community and extrapolate insights for enhancing the experiences and support of other novice creatives on social media platforms. We identify two key aspects -- \textbf{authenticity} and \textbf{growth} -- as pivotal areas for fostering a better environment for novice poets. A lack of support, could lead to a dearth of motivation engendering homogenized content production, wherein repetitive and generic content saturates the platform. Such a scenario will also impair the platform's ability to cater to the diverse preferences and expectations of its stakeholders, thereby curbing its potential for sustained growth and relevance. 

\subsubsection{Supporting authenticity}
 Authenticity on social media has been associated with \textit{being real} in a \textit{creative} manner \cite{uski2016}. Several researchers have positioned this as a social construct, that is defined by creators and viewers alike \cite{Cialdini1998SocialIS, barta}. In our study, the participants have expressed dissatisfaction with the homogeneity of content within the platform, both in terms of viewing other works and sharing their own. The creatives' preference lies in attaining more creative freedom and reduced constraints while engaging with the platform. To support this need, the main issue troubling the participants is the platform's preference for promoting content based solely on engagement. To combat that, Simpson and Semaan (2023) suggested platforms allow creatives to opt into desired audience sizes or choose whether to be categorized algorithmically, decoupling platform metrics from creative success \cite{simpson2023}. We extend several other ways to support creatives' authenticity and freedom.

%cite andalibi paper on tiktok authenticity
\textbf{Explaining visibility logic:}
For novices on Instagram, the call for heightened transparency in the algorithm is especially pertinent. Particularly poets, who predominantly work with textual content, encounter a unique struggle within an algorithm primarily designed for images. The existing perceived algorithmic bias toward visual elements poses a challenge for novice poets seeking to share their predominantly text-based creations.

To address this, a nuanced approach to transparency is essential. By shedding light on how the algorithm prioritizes and interacts with textual content, Instagram can better equip novice poets to navigate these dynamics. Similarly, exploring alternative explanations of the algorithm that address the specific needs of diverse content creators is also imperative. While constraints may persist with algorithms, transparency can offer clarity, reducing uncertainties linked to social media algorithms \cite{brooke2021sm}. 

\textbf{Flexibility in Choice:} To address participant preferences for authenticity, Instagram might consider introducing optional features for creators to underscore authenticity in their content. For example, poets could tag their posts as \textit{original} and users could filter their recommendations accordingly to either want \textit{personalized} or \textit{original} work. This enables the algorithm to better align with creatives' intentions and cater to users specifically seeking authentic creative work. 

%Lately, Instagram periodically asks its users if it likes certain content, and explains why certain content is shown on the feed. However, the nuance is still missing. Posts are still generic, and engagement is still the key driving factor. As \cite{simpson2023} suggested, differentiating metrics of success, such that there is separating of \textit{platform metrics and creative success} is necessary in this case. 

\textbf{Supporting community-based curation:} 
He et al., (2023), encouraged a \textit{democratic policy} that gives community leaders and members control over their recommended content \cite{he2023cura}. They introduced \textit{Cura}, a system with a recommendation algorithm specifically designed to recommend community-specific, focused content \cite{he2023cura}. Drawing from such research, and our findings,  we suggest that platforms encourage the use of user-generated, widely-adopted curation tags to establish a collective voice for community-based recommendations. While platforms like Instagram employ hashtags, uncertainties persist regarding their functionality as they are also algorithmically mediated. Along with specific community-based recommendation systems, enhancing user understanding and utilization of hashtags, in addition to algorithmic clarity, can empower users to participate in community-driven curation.

%rewrite above section
\subsubsection{Supporting growth}
As our findings suggest, which is consistent with prior studies, most new content creators are hobbyists, who with growth may or may not look for monetization opportunities \cite{Sharma2022ItsAB, wu2019}. However, with algorithmic mediation, novice creatives deal with constraints on their works, which in turn hinders growth and impacts their motivation for sharing on digital platforms. To enhance the growth of novice creators, we suggest that there is equity in visibility for novice creators.  

\textbf{Equity in visibility:}
We argue that the recommendation algorithm is sensitive to the unique needs faced by novice content creators, ignoring which can monopolise digital spaces \cite{10.1145/3555149}, and hinder the growth of new content creators. Factors such as account age, growth trajectory, and engagement trends should be considered in order to adjust recommendations accordingly. Researchers have discussed equity for monetizing creators, to help create a fairer space for all monetizing creatives regardless of sociocultural factors \cite{10.1145/3555149}. However, we encourage that platforms support growth opportunities for novices along with professionalized and monetizing creatives \cite{10.1145/2757226.2764774}. Therefore, implementing features to equally allow novices who do not have the experience, following, or resources of long-term users would help them grow, while reducing the uncertainties with the algorithm. 

%Currently, newcomers on Instagram are recommended content randomly during their first few uses, for the algorithm to know them better, in order to personalize their experiences. Likewise, we suggest transparency in implementing algorithmic interventions, wherein newcomers are allowed to understand how their profile is being recommended to others, along with clearer factors on which the process of curation can be based. 

%We believe that this focuses on equity in terms of visibility on digital platforms, wherein the recommendation systems have more nuance, such that growth for hobbyists and the proliferation of user-generated content is prioritized and supported as much as users who professionalise and monetize on their content. 

Ways to implement that can be, opt-in discovery features, such as dedicated sections or recommendation feeds specifically designed to highlight emerging poets or more broadly- new creatives. Meta claims to be creating tools to \textit{discover new voices} for novice audio artists \cite{fbIntroducingVoicebox} to particularly support them. A similar approach on Instagram could level the playing field for novices, and allow them to be discovered, further diversifying general content.

\section{Limitations and Future Work}

 This study is subject to several limitations. Firstly, participant bias is possible, as all interviewees exclusively composed poetry in English. The consideration of distinct linguistic dialects may necessitate separate studies, given the pivotal role of language in poetry \cite{harvardPoetryLanguages}.
Another limitation concerns potential incomplete information disclosure due to personal reasons among some interviewees \cite{pannucci2010identifying}. However, as the first author was a complete-member, we expect the disclosures to be honest and thorough.
Furthermore, Instagram's algorithm, the primary focus of this study, is dynamic and continually evolving. These ongoing algorithmic changes \cite{igback} may influence users' perceptions and behaviors along with data collection via hashtags because of potential bias due to the favoring of certain themes, perspectives, and user engagement patterns, thus limiting the representativeness of the collected data. Consequently, this underscores the importance of considering the evolving nature of the platform in future studies.

Future studies should attempt to provide more generalizable findings, by investigating other non-monetizing communities that have to be creative in order to meet their needs. Examples of this include grassroots organizations, DIY artists, and other hobbyist creatives. Furthermore, with consistent algorithmic changes, and creativity within the advent of newer technologies and digital realms, \textit{Algorithmic Mediated Creative Labor} requires a more dynamic investigation. Technological transience and its impacts were highlighted slightly in our study, as participants spoke of the precarious and confusing nature of algorithms (P13). Future studies should also consider \textit{technological change}, a crucial aspect when supporting less mainstream creators who may lack resources or knowledge to adapt.

\section{Conclusion}

Our research investigates Instagram poets, with a particular emphasis on their novice status, creative methodologies, value systems, as well as their comprehension and interpretation of the platform's underlying logic to extend work on novice creators' use of social media for content creation. To the best of our knowledge, this investigation represents the first in studying the poetry community and, by extension, exploring novice creative individuals who create without intentions of monetizing. We explore their interactions within this digital environment and their perceptions of the platform, unveiling distinctive insights unique to the poetry community. These insights underscore a dilemma faced by these creators – whether to remain authentic to their artistic identities or to make adjustments and conform to the platform's algorithms in order to gain visibility. We proceed to discuss the ramifications of the limited support provided to such novice and non-monetizing creators to define \textit{Algorithmically Mediated Creative Labor} as the additional labor required by non-monetizing creatives to sustain on platforms like Instagram. Our argument posits that platforms prioritize and extend support to such creators, given their pivotal role in generating UGC, and increasing diversity of platform content.

%\begin{acks}
%\end{acks}

\bibliographystyle{ACM-Reference-Format}
\bibliography{new-bib}

%\input{references}

\appendix

\end{document}